\documentclass{ws-ijmpa}

\begin{document}

\title{Prepotential approach to exact and quasi-exact
solvabilities of Hermitian and non-Hermitian Hamiltonians}

\author{Choon-Lin Ho}

\address{Department of Physics, Tamkang University,
Tamsui 251, Taiwan, Republic of China}

\date{Jan 7, 2008}
\maketitle 

\begin{abstract}

In this talk I present a simple and unified approach to both exact
and quasi-exact solvabilities of the one-dimensional Schr\"odinger
equation.  It is based on the prepotential together with Bethe
ansatz equations. This approach gives the potential as well as the
eigenfunctions and eigenvalues simultaneously. In this approach
the system is completely defined by the choice of the change of
variables, and the so-called zero-th order prepotential. We
illustrate the approach by several examples of Hermitian and
non-Hermitian Hamiltonians with real energies.  The method can be
easily extended to the constructions of exactly and quasi-exactly
solvable Dirac, Pauli, and Fokker-Planck equations, and to
quasinormal modes.

\end{abstract}



\section{Introduction}

I am greatly honored to be able to participate in this conference
celebrating Prof. Yang's 85th birthday.  We wish Prof. Yang a very
happy birthday, and many more to come.

In this talk I would like to present a constructive approach to
both exact and quasi-exact solvable one-dimensional Schr\"odinger
equations.  Everyone knows what exact solvability means.  It is
fair to say that all of us learned the principles of quantum
mechanics through several well-known exactly solvable (ES) models,
such as the infinite square well, the harmonic oscillator, the
hydrogen atom, etc. But in actual life, exactly solvable systems
are hard to come by. Most systems one encounters are non-solvable,
and one has to adopt various approximation schemes to solve them.

Two decades ago, a new class of potentials which are intermediate
to ES potentials and non solvable ones have been found for the
Schr\"odinger equation. These are called quasi-exactly solvable
(QES) models for which it is possible to determine algebraically a
part of the spectrum but not the whole spectrum
\cite{TU,Tur1,Tur,Ush,O1,Tur2}. The simplest QES model first
discovered is the sextic oscillator \cite{TU}. The discovery of
this class of spectral problems has greatly enlarged the number of
physical systems which we can study analytically.

Here I would like to discuss a novel approach that treats both ES
and QES systems on the same footing \cite{Ho07}.  It is based on
the so-called prepotential together with the Bethe ansatz
equations.

Since QES systems are less familiar to most people, I shall give a
very brief introduction to QES theory before the prepotential
approach is discussed.

\section{A brief introduction to QES theory}

The essence of quasi-exact solvability is most easily explained as
follows \cite{Ush}.  It is well known that any Hamiltonian $H$ can
be represented as an infinite-dimensional Hermitian matrix
\begin{eqnarray}
H=\left[
\begin{array}{ccccc}
H_{00} & H_{01}       & \cdots     & H_{0N}   &\cdots   \\
H_{10} & H_{11}       & \cdots     & H_{1N}   &\cdots   \\
\cdots &\cdots   &\cdots  &\cdots  &\cdots \\
H_{N0} & H_{N1}       & \cdots     & H_{NN}   &\cdots   \\
\cdots &\cdots   &\cdots  &\cdots  &\cdots
\end{array}
\right],
\end{eqnarray}
where the elements $H_{nm}=\langle\psi_n|H|\psi_m\rangle$ depend
on the choice of a complete set of orthonormal functions $\psi_n$
forming a basis of the Hilbert space.  The solution of the
spectral problem then reduces to a diagonalization of matrix
$\{H_{nm}\}$. Unfortunately, contrary to the case of finite
matrices, there is no general algebraic rules that would allow one
to diagonalize the infinite-dimensional $H$.

If the matrix $\{H_{nm}\}$ is very specific so that it can be
reduced to the diagonal form with the aid of an algebraic process,
then  the quantal system is exactly solvable.
\begin{eqnarray}
H=\left[
\begin{array}{cccccc}
H_{00} & 0       & 0     & 0      & 0 &\cdots   \\
0      & H_{11}  & 0     & 0      & 0 &\cdots  \\
0      & 0       & H_{22}& 0      & 0 &\cdots  \\
0      & 0       & 0     & H_{33} & 0 &\cdots  \\
\cdots &\cdots   &\cdots  &\cdots  &\cdots &\cdots
\end{array}
\right].
\end{eqnarray}
 The most well-known
example is the harmonic oscillator.

Suppose a Hamiltonian $H$ is reducible to a block form
\begin{eqnarray}
H=\left[
\begin{array}{cccccccc}
H_{00} & H_{01} &\cdots
&H_{0N} & 0 & 0 &\cdots & 0   \\
H_{10} &H_{11} &\cdots
 &H_{1N}& 0 & 0 &\dots  & 0  \\
\cdots &\cdots &\cdots&\cdots&\cdots &\cdots &\cdots
&\cdots   \\
H_{N0}& H_{N1}&\cdots
&H_{NN} & 0 & 0 &\dots  & 0  \\
0 & 0 & \cdots &0 &*\,\,  &*\,\,  &\cdots &\cdots\\
0 & 0 & \cdots &0 &*\,\,  &*\,\,  &\cdots  &\cdots\\
\cdots &\cdots   &\cdots  &\cdots &*\,\,  &*\,\,  &\cdots &\cdots\\
0 & 0 & \cdots &0 &*\,\,  &*\,\,  &\cdots &\cdots\\
\cdots &\cdots   &\cdots  &\cdots &*\,\,  &*\,\,  &\cdots &\cdots
\end{array}
\right],
\end{eqnarray}
where the block in the upper left corner is an $(N+1)\times (N+1)$
Hermitian matrix, and the block  with asterisks is an infinite
matrix with non-vanishing elements.  The upper left block can be
diagonalized without touching the infinite one.  This means that
one can determine only a part of the spectrum of $H$ with $N+1$
eigenvalues and eigenfunctions.  This system is called QES.

QES models are not simply mathematical constructs. Although the
first and simplest example of QES model is the sextic oscillator,
physical QES systems have been found.  For examples, the system of
two electrons in a three-dimensional oscillator potential
\cite{SG}, two-dimensional charged particle in Coulomb and uniform
magnetic fields (the Schr\"odinger, Klein-Gordon, and Dirac cases
are discussed in Refs. \refcite{Taut}, \refcite{VP} and
\refcite{Ho1}, respectively), and electron on a lattice in
magnetic field (Azbel-Hofstadter problem)
 \cite{WZ}.

Usually a QES problem admits a certain underlying Lie algebraic
symmetry which is responsible for the quasi-exact solutions.  Such
underlying symmetry is most easily studied in the Lie-algebraic
approach \cite{Tur,O1}.  The essence of this approach is as
follows. Consider a Schr\"odinger equation $H\psi=E\psi$ with
Hamiltonian $H=-d^2/dx^2 + V(x)$ and wave function $\phi (x)$.
Here $x$ belongs either to the interval $(-\infty,\infty)$ or
$[0,\infty)$. Now suppose we make an ``imaginary gauge
transformation" on the function $\phi$: $\phi (x)= {\tilde
\phi}(x) e^{-W_0(x)}$.  For physical systems which we are
interested in, the phase factor $\exp(-W_0(x))$ is responsible for
the asymptotic behaviors of the wave function so as to ensure
normalizability. The function ${\tilde\phi}(x)$ satisfies a
Schr\"odinger equation with a gauge transformed Hamiltonian
$H_W=e^{W_0} H e^{-W_0}$. Suppose $H_W$ can be written as a
quadratic combination of the generators $J^a$ of some Lie algebra
with a finite dimensional representation, i.e.,
\begin{eqnarray}
H_W=\sum c_{ab}J^a J^b + \sum c_a J^a + {\rm real\ constant},
\end{eqnarray}
where the constants $c_{ab}$ and $c_a$ are usually taken to be
real \cite{Tur,O1,Tur2}. Then within this finite dimensional
Hilbert space the Hamiltonian $H_W$ can be diagonalized, and
therefore a finite number of eigenstates are solvable.  Hence the
system described by $H$ is  QES. For one-dimensional QES systems
the most general Lie algebra is $sl(2)$.  The generators $J^a$ of
the $sl(2)$ Lie algebra  take the differential forms: $ J^+ = z^2
d_z - nz~,~ J^0=z d_z-n/2~,~J^-=d_z$ ($n=0,1,2,\ldots$). The
variables $x$ and $z$ are related by some function. $n$ is the
degree of the eigenfunctions $\tilde\phi$, which are polynomials
in a $(n+1)$-dimensional Hilbert space with the basis $\langle
1,z,z^2,\ldots,z^n\rangle$.

The Lie-algebraic approach to QES models excels in revealing the
underlying symmetry of a QES system explicitly.  However,
solutions of QES states are more directly found in the analytic
approach based on the Bethe ansatz equations \cite{Ush}.  In this
analytic approach the form of the wave functions containing some
parameters are assumed from the very beginning, and these
parameters are fitted to make the ansatz compatible with the
potential under consideration .

Recently, a different approach to QES models has emerged
\cite{Ho,ST}. In this approach the emphasis shifts from the
potential to the so-called prepotential (or superpotential), a
concept which plays a fundamental role in supersymmetric quantum
mechanics (we note here that the prepotential considered here is
the integral of the superpotetial in Refs.~\refcite{Ho} and
\refcite{Cooper}). Prepotential has been extensively employed to
study classical and quantum integrability in Calogero-Moser
systems \cite{CS}.  The merit of this approach is that the form of
the potential of the system concerned needs not be assumed from
the beginning.  All information about the system is contained in
the prepotential and the solutions, or roots, of the Bethe ansatz
equations (in terms of certain transformed coordinates). The
prepotential and the roots determine the potential as well as the
eigenfunctions and eigenvalues simultaneously. Also, in this
approach exact and quasi-exact solvabilities can be treated on the
same footing. Furthermore, such approach facilitates extension of
the QES theory from the Schr\"odinger equation to equations for
multi-component wave functions \cite{Ho,B}. More recently, QES
theory was extended to the Fokker-Planck equations also via the
prepotential approach \cite{HS}.

The emphasis of the works just mentioned was placed on the
feasibility and elegance of the prepotential approach.  But in
these works the forms of the prepotential and the required change
of coordinates were either directly adapted from the known ES and
QES models, or given as known for the new QES systems.  To make
the prepotential approach a satisfactory approach one must be able
to determine the choice of coordinate transformation and the
prepotential from the very beginning.  This was recently achieved
for certain classes of coordinates and prepotentials \cite{Ho07}.
In what follows we shall present the main ideas of this approach,
and demonstrate how the simplest ES and QES models could be easily
derived.

\section{Prepotential approach}

Suppose $\phi_0(x)~(-\infty<x<\infty))$ is the ground state, with
zero energy, of a Hamiltonian $H_0$: $H_0\phi_0=0$. By the well
known oscillation theorem $\phi_0$ is nodeless, and thus can be
written as $\phi_0\equiv e^{-W_0(x)}$, where $W_0(x)$ is a regular
function of $x$. For the square-integrable $\phi_0$, this is the
simplest example of quasi-exact solvability. This implies that the
potential $V_0$ is completely determined by $W_0$:
$V_0={W_0^\prime}^2 - W_0^{\prime\prime}$, and consequently, the
Hamiltonian is factorizable (we adopt the unit system in which
$\hbar$ and the mass $m$ of the particle are such that
$\hbar=2m=1$):
\begin{equation}
H_0=\left(-\frac{d}{dx}+W_0^\prime\right)\left(\frac{d}{dx}+W_0^\prime\right).
     \label{facHam}
\end{equation}
We shall call $W_0(x)$ the zero-th order prepotential.

Consider now a wave function $\phi_N$ ($N\geq 0$) which is related
to $\phi_0$ of $H_0$ by $\phi_N=\phi_0{\tilde \phi}_N$, where
\begin{eqnarray}
 {\tilde\phi}_N=(z-z_1)(z-z_2)\cdots
(z-z_N),~~{\tilde\phi}_0\equiv 1. \label{phi-2}
\end{eqnarray}
 Here $z=z(x)$ is some function of $x$. In taking the form of
${\tilde\phi}_N$ in Eq.~(\ref{phi-2}) I have assumed that the only
singularities of the system are $z=\pm\infty$.  Other situations
are discussed in Ref.~\refcite{Ho07}. The function
${\tilde\phi}_N$ is a polynomial in an $(N+1)$-dimensional Hilbert
space with the basis $\langle 1,z,z^2,\ldots,z^N \rangle$.  One
can rewrite $\phi_N$ as
\begin{eqnarray}
 \phi_N =\exp\left(- W_N(x,\{z_k\})
\right), \label{f2}
\end{eqnarray}
with the $N$-th order prepotential $W_N$ being defined by
\begin{eqnarray}
W_N(x,\{z_k\}) = W_0(x) - \sum_{k=1}^N \ln |z(x)-z_k|. \label{W}
\end{eqnarray}
Operating on $\phi_N$ by the operator $-d^2/dx^2$ results in a
Schr\"odinger equation $H_N\phi_N=0$, where
\begin{eqnarray}
H_N &=&-\frac{d^2}{dx^2} + V_N,\\
V_N&\equiv&  W_N^{\prime 2} - W_N^{\prime\prime}.
\end{eqnarray}
From Eq.~(\ref{W}), $V_N$ has the form $V_N=V_0+\Delta V_N$, where
$V_0=W_0^{\prime 2} - W_0^{\prime\prime}$, and
\begin{eqnarray}
\Delta V_N \equiv -2\left(W_0^\prime z^\prime
-\frac{z^{\prime\prime}}{2}\right)\sum_{k=1}^N \frac{1}{z-z_k} +
\sum_{k,l\atop k\neq l} \frac{z^{\prime 2}}{(z-z_k)(z-z_l)}.
\label{dV}
\end{eqnarray}
Here the prime denotes derivative w.r.t. the variable $x$.

$\Delta V_N$ is generally a meromorphic function of $z$ with at
most simple poles. Let us demand that the residues of the simple
poles, $z_{k}$, $k=1,\ldots, N$ should all vanish. This will
result in a set of algebraic equations which the parameters
$\{z_k\}$ must satisfy. These equations are called the Bethe
ansatz equations for $\{z_k\}$.  With $\{z_k\}$ satisfying the
Bethe ansatz equations, $\Delta V_N$ will have no simple poles at
$\{z_k\}$ but it still generally depends on $\{z_k\}$. Thus the
form of $V_N$ is determined by the choice of $z^{\prime 2}$ and
$W_0(x)$.

In what follows we would like to demonstrate that the choice of
$z^{\prime 2}$ and $W_0^\prime z^\prime$ determine the nature of
solvability of the quantal system.   We shall restrict our
consideration only to those cases where $W_0^\prime z^\prime=P_m
(z)$, $z^{\prime 2}=Q_n(z)$ and $z^{\prime\prime}=R_l(z)$ are
polynomials in $z$ of degree $m$, $n$ and $l$, respectively.  Of
course, $Q_n(z)$ and $R_l(z)$ are not independent.  In fact, from
$2z^{\prime\prime}=dz^{\prime 2}/dz$ we have $l=n-1$ and
\begin{eqnarray}
z^{\prime\prime}=R_{n-1}(z)=\frac{1}{2}\frac{dQ_n(z)}{dz}.
\end{eqnarray}
Consequently, the variables $x$ and $z$ are related by (we assume
$z(x)$ is invertible for practical purposes)
\begin{eqnarray}
x(z)=\pm \int^z \frac{dz}{\sqrt{Q_n(z)}},\label{z(x)}
\end{eqnarray}
and the prepotential $W_0(x)$ is determined as
\begin{eqnarray}
W_0(x)=\int^x dx
\left(\frac{P_m(z)}{\sqrt{Q_n(z)}}\right)_{z=z(x)}.\label{W0}
\end{eqnarray}
Eqs.~(\ref{z(x)}) and (\ref{W0}) define the transformation $z(x)$
and the corresponding prepotential $W_0(x)$. Thus, $P_m(z)$ and
$Q_n(z)$ determine the quantum system.  Of course, the choice of
$P_m$ and $Q_n$ must be such that $W_0$ derived from (\ref{W0})
must ensure normalizability of $\phi_0=\exp(-W_0)$.

Now depending on the degrees of the polynomials $P_m$ and $Q_n$,
we have the following situations:

\begin{enumerate}
\item[(i)] if $\max\{m,n-1\}\leq 1$, then in $V_N(x)$
the parameter $N$ and the roots $z_k$'s will only appear as an
additive constant and not in any term involving powers of $z$.
Such system is then exactly solvable;

\item[(ii)] if $\max\{m,n-1\}=2$, then $N$ will appear in
the first power term in $z$, but $z_k$'s only in an additive term.
This system then belongs to the so-called type 1 QES system
defined in Ref.~\refcite{Tur}, i.e., for each $N\geq 0$, $V_N$
admits $N+1$ solvable states with the eigenvalues being given by
the $N+1$ sets of roots $z_k$'s.  This is the main type of QES
systems considered in the literature;

\item[(iii)] if $\min\{m,n-1\}\geq 3$, then not only $N$ but also
$z_k$'s will appear in terms involving powers of $z$. This means
that for each $N\geq 0$, there are $N+1$ different potentials
$V_N$, differing in several parameters in terms involving powers
of $z$, have the same eigenvalue (when the additive constant, or
the zero point, is appropriately adjusted). When $z_k$'s appear
only in the first power term in $z$, such systems are called type
2 QES systems in Ref.~\refcite{Tur}.  We see that QES models of
higher types are possible.
\end{enumerate}

These general situations are considered for cases where $z^{\prime
2}=Q_2(z)$ are quadratic in $z$ (this includes $Q_1$ and $Q_0$ as
special cases if some of the coefficients vanish) in
Ref.~\refcite{Ho07}. This choice of $z^{\prime 2}$ covers most of
the known ES shape-invariant potentials in Ref.~\refcite{Cooper}
and the $sl(2)$-based QES systems in Ref.~\refcite{Tur}, and a new
one discussed in Refs.~\refcite{ST} and \refcite{HS}. Such
coordinates are called ``sinusoidal coordinates", which include
quadratic polynomials, trigonometric, hyperbolic, and exponential
types. The connection of sinusoidal coordinates with ES theory has
been extensively discussed in Ref.~\refcite{OS}.  For simplicity,
in this talk I shall only discuss cases in which $z^{\prime 2}$ is
linear in $z$, i.e., $z^{\prime 2}=Q_1(z)=q_1z+q_0$, where $q_1$
and $q_0$ are real constants.

With this choice of $z^{\prime 2}$, we have
\begin{eqnarray}
 V_N ={W_0^\prime}^2 - W_0^{\prime\prime} -2\sum_{k=1}^N
\frac{1}{z-z_k}\left\{P_m(z) - \frac{q_1}{4} - \sum_{l\neq k}
\frac{Q_1(z_k)}{z_k-z_l}\right\}. \label{dV1}
\end{eqnarray}
In deriving Eq.~(\ref{dV1}) use has been made of the following
identities:
\begin{eqnarray}
\sum_{k,l=1\atop k\neq l}^N \frac{1}{(z-z_k)(z-z_l)} &=& 2
\sum_{k,l=1\atop k\neq l}^N
\frac{1}{z-z_k}\left(\frac{1}{z_k-z_l}\right),\\
\sum_{k,l=1\atop k\neq l}^N \frac{z}{(z-z_k)(z-z_l)} &=& 2
\sum_{k,l=1\atop k\neq l}^N
\frac{1}{z-z_k}\left(\frac{z_k}{z_k-z_l}\right).
\end{eqnarray}
 Demanding the residues at $z_k$'s vanish gives the set
of Bethe ansatz equations
\begin{eqnarray}
P_m(z_k) - \frac{q_1}{4} - \sum_{l\neq k}
\frac{Q_1(z_k)}{z_k-z_l}=0,~~k=1,2,\ldots,N. \label{BAE}
\end{eqnarray}
Putting back the set of roots $z_k$ into Eq.~(\ref{dV1}), we
obtain a potential $V_N(x)$ without simple poles. The degree of
$P_m(z)$ determines the nature of the solvability of the system,
namely, for $m=1,2,3,\ldots$, the system is, respectively, ES,
type 1 QES, type 2 QES, and higher types QES, as discussed
generally in the last section.

To proceed further, we must specify $P_m(z)$.  Below I shall
discuss cases with $m\leq 2$, i.e., only ES and type 1 QES models.

\section{Examples}

Let us take $Q_1(z)=4Az +q_0$ ($q_1=4A$), where $A,~q_0$ are real
constants. This implies
 $z^{\prime\prime}(x)=R_0(z)=2A$.  Hence, the general solution of $z(x)$
is a quadratic form of $x$:
\begin{eqnarray}
z(x)=Ax^2 + Bx + C,~~A,~B,~C {\rm : real\  constants}. \label{z}
\end{eqnarray}
It is easily checked that $q_0$ is related to $A,~B$ and $C$
through $z^{\prime 2}=4Az +B^2-4AC$.

We now show how some ES and QES models can be constructed in the
prepotential approach by taking different values of $m$.

\subsection{Exactly solvable cases: $m=1$}

Suppose $m=1$ so that $P_1(z)=A_1z + A_0$ ($A_1,~A_0$ real).  By
writing $P_1=A_1(z-z_k)+A_1z_k+A_0$, one obtains from
Eq.~(\ref{dV1}) that
\begin{eqnarray}
\Delta V_N &=& -2A_1\sum_{k=1}^N 1 -
2\sum_{k=1}^N\frac{1}{z-z_k}\left\{P_1(z_k) -A -
\sum_{l\neq k} \frac{Q_1(z_k)}{z_k-z_l}\right\}\nonumber\\
&=&-2A_1N. \label{dV-case1}
\end{eqnarray}
The last term in braces in Eq.~(\ref{dV-case1}) vanishes when
$z_k$'s satisfy the Bethe ansatz equations (\ref{BAE}).   Now $N$
only appears as a parameter in an additive term, and not in terms
involving powers of $z$ in $\Delta V_N$. The roots $z_k$'s do not
appear at all. The additive term can be treated as the eigenvalue.
The Schr\"odinger equation reads
\begin{eqnarray}
\left(-\frac{d^2}{dx^2} + W_0^{\prime
2}-W_0^{\prime\prime}\right)e^{-W_N}=2A_1Ne^{-W_N}.
\end{eqnarray}
We see that the potential $W_0^{\prime 2}-W_0^{\prime\prime}$ is
ES: by varying $N$, one obtains all the eigenvalues $2A_1N$ and
the eigenfunctions $\phi_N=\exp(-W_N)$.

As an example, let us take $Q_1(z)=1$ and $P_1(z)=bz$. A
particular solution of $z$ is $z(x)=x$. From Eq.~(\ref{W0}) one
gets
\begin{eqnarray}
W_0(x)=\int^x \frac{bx}{1}dx=\frac{bx^2}{2}+{\rm const.}
\end{eqnarray}
The integration constant is to be determined by normalization of
the wave function.  For $\phi_0=\exp(-W_0)$ to be
square-integrable, one must assume $b>0$. The BAE (\ref{BAE}) are:
\begin{eqnarray}
bx_k-\sum_{l\neq k} \frac{1}{x_k-x_l}=0, ~~k=1,\ldots,N,
\label{BAE-Osc}
\end{eqnarray}
and the potential is
\begin{eqnarray}
V_0=b^2 x^2 -b,~~\Delta V_N=-2Nb.
\end{eqnarray}
This leads to the Schr\"odinger equation:
\begin{eqnarray}
\left(-\frac{d^2}{dx^2} + b^2 x^2 \right)e^{-W_N}=b(2N+1)e^{-W_N}.
\end{eqnarray}
  This system is just the well-known simple
harmonic oscillator.

We note here that by rescaling $\sqrt{b}x_k\to x_k$,
Eq.~(\ref{BAE-Osc}) will have $b=1$.   The resulted equations are
the equations that determine the zeros of the Hermite polynomials
$H_N(x)$ as found by Stieltjes \cite{Stiel,Szego,CS}. Hence we
have reproduced the well known wave functions for the harmonic
oscillator, namely, $\phi_N=\exp(-W_N)\sim
\exp(-bx^2/2)H_N(\sqrt{b}x)$.

\subsection{Type 1 quasi-exactly solvable cases: $m=2$}

Next we consider $P_2(z)=A_2 z^2 + A_1 z+ A_0$.  By a similar
argument we obtain
\begin{eqnarray}
\Delta V_N= -2A_2 N z -2A_2\sum_{k=1}^N z_k -2A_1 N.
\end{eqnarray}
The Schr\"odinger can be written as
\begin{eqnarray}
\left(-\frac{d^2}{dx^2} + W_0^{\prime 2}-W_0^{\prime\prime}-
2A_2Nz \right)e^{-W_N}=2\left(A_2\sum_{k=1}^N z_k+
A_1N\right)e^{-W_N}. \label{SE-case2}
\end{eqnarray}
Unlike the previous case, now $N$ not only appears in an additive
constant term but also in the term with $z$, and the set of roots
$z_k$'s appear in the additive term. This system is the so-called
type 1 QES models.  Type 1 QES models classified as class VI in
Ref.~\refcite{Tur} belong to this category.

A well-known example is the sextic oscillator, the simplest QES
model of this type \cite{TU}, which is termed as class VI in
Ref.~\refcite{Tur}. In our prepotential approach, this system is
defined by $z(x)=x^2$ and $P_2(z)=2(az^2 + bz)$.
 Then
$Q_1(z)=4z$, and
\begin{eqnarray}
W_0(x)=\int^x \frac{ax^4 + bx^2}{\sqrt{x^2}}dx=\frac{1}{4}ax^4+
\frac{1}{2}bx^2 + {\rm const.}
\end{eqnarray}
Here $a>0$ to ensure square-integrability of the wave function.
The BAE (\ref{BAE}) are:
\begin{eqnarray}
2az_k^2 +2bz_k -1 - 4\sum_{l\neq k}\frac{z_k}{z_k-z_l} =0,~~~
k=1,\ldots,N,
\end{eqnarray}
and the potential is
\begin{eqnarray}
V_N=a^2 x^6 +2ab x^4+ \left[b^2-\left(4N +3\right)a\right]x^2
-4a\sum_k z_k -(4N+1)b.
\end{eqnarray}
This is just the well-known QES sextic oscillator first discovered
in Ref.~\refcite{TU}.

These two examples show how simply and directly the two most
classic representatives of ES and QES models are derived in the
prepotential approach.   Other ES and QES models can be
constructed in a similar way.  Particularly, we can consider cases
where $z^{\prime 2}(x)=Q_2(z)$ with $q_2\neq 0$. Here $z(x)$ is
again some sinusoidal coordinates, which include the exponential,
hyperbolic and trigonometric functions.  By taking appropriate
$P_m(z)$, one can reconstruct class I, II and X QES models in
Ref.~\refcite{Tur}, and some of the ES models listed in
Ref.~\refcite{Cooper}, namely, the Morse potential, the Scarf I
and II potentials, and the P\"oschl-Teller potential \cite{Ho07}.

\section{Non-Hermitian potentials with real spectra}

Now we turn to a different kind of spectral problem.  About ten
years ago, it was noted that with properly defined boundary
conditions the spectrum of the non-Hermitian Hamiltonian
$H=p^2+x^2(ix)^\nu$ ($\nu\geq 0$) is real and positive \cite{BB1}.
It is believed that the reality of the spectrum of this
Hamiltonian is a consequence of its unbroken $\cal{PT}$ symmetry.
This discovery immediately sparked great interest in searching for
new systems with such properties, and in understanding the reasons
behind the reality of their eigenvalues (for review, see e.g.
Ref.~\refcite{B}).

Soon after the appearance of Ref.~\refcite{BB1}, it was found
that, by allowing non-Hermitian $\cal{PT}$-symmetric Hamiltonians,
a QES polynomial potential can be quartic in its variable
\cite{BB2}. Previously, it was believed the lowest-degree
one-dimensional QES polynomial potential is sextic, namely, that
presented in Sect.~4. Now let us demonstrate how simply the
Hamiltonian discovered in Ref.~\refcite{BB2} is derived in our
approach if we allow $P_m(z)$, or equivalently $W_0(x)$, to be
complex.

\subsection{$\cal{PT}$-symmetric quartic case}

We take $z=x$ (i.e. $Q_0=1$), and $P_2(x)=a x^2 + b x + c$ for
type 1 QES system.  Eq.~(\ref{W0}) then gives $W_0(x)=ax^3/3 +
bx^2/2 + cx$.  It is obvious that, with $a,b$ and $c$ real, the
wave functions are not normalizable as the factor $\exp(-W_0)$
does not converge on the whole line. Hence a QES Hamiltonian with
quartic polynomial potential is impossible with real coefficients
in $W_0$.

What if we allow $P_2(x)$ to be complex?  Suppose we take
$a=i\alpha,~b=\beta$ and $c=i\gamma$, where $\alpha,~\beta>0$ and
$\gamma$ are real.   Obviously $\exp(-W_0)$ is now
square-integrable on the whole line for $b>0$. From
Eqs.~(\ref{dV1}) and (\ref{BAE}) we obtain straightforwardly the
type 1 QES potential
\begin{eqnarray}
V_N(x)&=&-\alpha^2 x^4 + 2i\alpha\beta x^3 + \left(\beta^2
-2\alpha\gamma\right)x^2
-2i\left[(N+1)\alpha-\beta\gamma\right]x\nonumber\\
&& - \left[\gamma^2 +\beta+2i\alpha\sum_k x_k\right],
\label{complex1}
\end{eqnarray}
with the BAE
\begin{eqnarray}
i\alpha x_k^2+\beta x_k + i\gamma - \sum_{l\neq k}
\frac{1}{x_k-x_l}=0,~~k=1,2,\ldots,N.
 \label{BAE-complex1}
\end{eqnarray}
$V_N$ given by Eq.~(\ref{complex1}) is precisely the
$\cal{PT}$-symmetric QES quartic potential obtained in
Ref.~\refcite{BB2} (in the notation of Ref.~\refcite{BB2}, the
potential has parameters $\alpha=1$, $\beta=a$, $\gamma=b$, and
$N+1=J$). In Ref.~\refcite{BB2} it was shown that this
$\cal{PT}$-symmetric Hamiltonian is also related to the
Lie-algebra $sl(2,C)$.   It is nice to see that in our
prepotential approach this potential can be so simply and directly
derived, without any knowledge of its underlying symmetries.

The properties of this system have been fully described in
Ref.~\refcite{BB2}, so we will not attempt a full discussion of
it. Let us just consider the two simplest cases with $N=0$ and
$N=1$.

For $N=0$, the potential is just $V_0$ with only one solvable
state corresponding to energy $E=0$.  For $N=1$, there are two
solvable states.  From Eq.~(\ref{complex1}) we can take the
potential to be
\begin{equation}
V_1=-\alpha^2 x^4 + 2i\alpha\beta x^3 + \left(\beta^2
-2\alpha\gamma\right)x^2 -2i(2\alpha-\beta\gamma)x
\end{equation}
with eigenvalues $\gamma^2 +\beta+2i\alpha x_1$.  The roots $x_1$
satisfy the BAE (\ref{BAE-complex1}): $i\alpha x_1^2+\beta x_1 +
i\gamma=0$. The solutions are: $x_1=i(\beta\pm
\sqrt{\beta^2+4\alpha\gamma})/2\alpha$.  Hence we see that for
$\beta^2+4\alpha\gamma >0$ the non-Hermitian potential $V_1$ has
two normalizable states with real energies $\gamma^2\pm
\sqrt{\beta^2+4\alpha\gamma}$.  But if $\beta^2+4\alpha\gamma <0$,
the energies are complex.  This is consistent with the numerical
results in Ref.~\refcite{BB2}.  For higher values of $N$ the
analysis is more complicated and numerics are needed \cite{BB2}.

\subsection{Sextic case without $\cal{PT}$ symmetry}

We now consider a second non-Hermitian QES system. Suppose we
assume the parameter $b$ in the sextic oscillator (class VI in
Ref.~\refcite{Tur}) in Sect.~4.2 to be purely imaginary: $b=
i\beta$, where $\beta$ and $a>0$ are real. The factor $\exp(-W_0)$
is square-integrable. Now the potential and BAE are:
\begin{eqnarray}
V_N&=&a^2 x^6 +2ia\beta x^4 - \left[\beta^2 + \left(4N
+3\right)a\right]x^2 -4a\sum_k z_k -i(4N+1)\beta,
\label{complex2}\\
&&2az_k^2 +2i\beta z_k -1 - 4\sum_{l\neq k}\frac{z_k}{z_k-z_l}
=0,~~~ k=1,\ldots,N. \label{BAE-complex2}
\end{eqnarray}
Again, this non-Hermitian system could have real eigenvalues.
Trivially for $N=0$ the potential $V_0$ admits one normalizable
eigenstate with real energy $E=0$.  For $N=2$, the solutions of
Eq.~(\ref{BAE-complex2}) are: $z_1=(-i\beta\pm
\sqrt{2a-\beta^2})/2a$. Then Eq.~(\ref{complex2}) becomes
\begin{eqnarray}
V_1=a^2 x^6 +2ia\beta x^4 - (\beta^2 + 7a)x^2 -3i\beta \pm
2\sqrt{2a-\beta^2}.\label{complex3}
\end{eqnarray}
Hence the non-Hermitian potential defined by the first four terms
in Eq.~(\ref{complex3}) has two real eigenvalues $E=\pm
2\sqrt{2a-\beta^2}$ if $\beta^2<2a$.  The energies become complex
if $\beta^2 > 2a$.  Like the quartic case just discussed, for
higher values of $N$ the analysis of its spectrum requires the aid
 of numerics.  Let us emphasize that this systems is not
 $\cal{PT}$-symmetric.

We note here that the same analysis can be applied to class I and
VII QES systems in Ref.~\refcite{Tur} to get other non-Hermitian
QES systems with real energies.

\section{Summary}

I hope I had given you a general idea of how simple and elegant
the prepotential approach to ES and QES models is.  In this talk I
have limited myself to discussing our approach only to systems
defined on the full line owing to time limitation.  Our approach
has been extended to ES and QES systems defined on half-line and
on finite intervals \cite{Ho07}. It can be easily extended to the
construction of exactly and quasi-exactly solvable Dirac, Pauli
\cite{Ho}, Fokker-Planck equations \cite{HS}, and to quasinormal
modes, which are eigen-modes with complex energies of a Hermitian
Hamiltonian \cite{CH}.

Of course, what I have presented here is quite preliminary.  Our
approach could be further developed. Firstly, so far we have only
considered cases for which $W_0^\prime z^\prime$ and $z^{\prime
2}$ are polynomials in $z$. It is interesting to consider other
possibilities for these two defining functions. Secondly, it is
extremely interesting to see how our formalism could be extended
to many-body systems, such as the renowned
Calogero-Sutherland-Moser systems.  The analogues of Bethe-ansatz
equations for these systems are not known yet. Lastly, although
some non-Hermitian Hamiltonians can be generated in this approach
by making some coefficients in $P_m(z)$ complex, as we have shown
here, a full treatment of these kind of systems usually requires
one to extend the basic variable $x$ to the complex plane
\cite{BB1,B,BB2}. Hence a better understanding of the prepotential
approach to non-Hermitian Hamiltonians is needed.  We hope to
report on our progress in these directions in the near future.

\section*{Acknowledgments}

I thank R. Sasaki for discussions and for bringing my attention to
sinusoidal coordinates.  I also thank A. Turbiner for
encouragement and stimulating comments on many-body systems. This
work is supported in part by the National Science Council (NSC) of
the Republic of China under Grant Nos. NSC 96-2112-M-032-007-MY3
and NSC 95-2911-M-032-001-MY2.




\end{document}